\documentclass[a4paper,11pt]{article}
\usepackage[dvipdfmx]{graphicx}
\usepackage{jinstpub} % for details on the use of the package, please see the JINST-author-manual
\usepackage{comment}

\title{\boldmath Development of an imager with high time resolution optical photon counter}

\author[a]{A. Sato}
\author[a]{T. Nakamori}
\author[b]{M. Shoji}
\author[a]{T. Sato}
\author[a,c]{K. Hashiyama}
\author[a]{M. Hasebe}
\author[a]{M. Maeshiro}
\author[a]{R. Sato}
\author[b]{R. Honda}
\author[b]{M. Miyahara}

\affiliation[a]{Yamagata University,\\1-4-12 Kojirakawa-machi, Yamagata 990-8560, Japan}
\affiliation[b]{KEK,\\1-1 Oho, Tsukuba, Ibaraki 305-0801, Japan}
\affiliation[c]{University of Tokyo,\\7-3-1 Hongo, Bunkyo, Tokyo 113-8654, Japan}

\emailAdd{s232623m@st.yamagata-u.ac.jp, nakamori@sci.kj.yamagata-u.ac.jp}

\abstract{
Astrophysical transient phenomena on sub-millisecond timescales,
such as fast radio bursts and giant radio pulses from the Crab pulsar,
have been primarily observed in radio wavebands.
To investigate their origins, a photon detector with high sensitivity and high time resolution is required also in other wavelengths.
Recently, we developed the Imager of MPPC-based Optical photoN counter from Yamagata (IMONY),
an observation system utilizing a Geiger-mode avalanche photodiode (GAPD) as a sensor.
The sensor consists of 64 pixels, each comprising a GAPD and a quenching resistor,
with pixel sizes of 75, 100, 150, and 200\,$\mu$m. 
Each pixel signal is read out independently, enabling single-photon detection.
After successfully observing the Crab pulsar using two Japanese telescopes,
we upgraded the readout boards to achieve a more compact and stable system.
The new system incorporates an analog application-specific integrated circuit (ASIC)
developed at KEK for multi-purpose fast readout for silicon photomultipliers.
This ASIC features a fast transimpedance amplifier and a comparator, independently processing 16 channels.
A Global Navigation Satellite System receiver and a Field Programmable Gate Array (FPGA)
provide timestamps for each detected photon with a resolution of 100 ns.
The FPGA transmits the acquired data to a PC via Ethernet.
This paper presents the details of the new system and the results of its initial evaluation.
}

\keywords{Detectors for UV, visible and IR photons; Photometers; Front-end electronics for detector readout; Optical detector readout concepts}

\begin{document}
\maketitle
\flushbottom

\section{Introduction}
\label{sec:intro}

The primary scientific objective of this study is to observe fast optical phenomena.
A key target is the enhancement of optical emission coinciding with giant radio pulses (GRPs) from the Crab pulsar.
The Crab pulsar is a rapidly rotating neutron star 
with a rotation period of approximately 34\,ms.
GRPs are transient radio bursts observed in the Crab pulsar, 
characterized by these sudden radio emissions on timescales ranging from ns to $\mu$s \cite{Hankins}.
These pulses are sometimes several hundred times stronger or more 
than the fluctuations in the background pulsar wind nebula (PWN) radiation.
Previous studies \cite{Strader} have reported that optical emission is enhanced by approximately 3\% in coincidence with GRPs.
However, the emission mechanism of GRPs remains unclear, and further multi-wavelength observations are required to understand their origin.
To address this situation, we have developed IMONY, a high-sensitivity, high-time-resolution optical photon detector \cite{Nak, Nak2, Has}.

This paper focuses on the system integration and laboratory evaluation of this newly developed instrument.  Similarities and differences in detector design and performance compared to other photon-counting systems are discussed in Ref.~\cite{Nak, Nak2}.

\section{IMONY}
\label{sec:imony}

We have developed the Imager of MPPC-based Optical photoN counter from Yamagata (IMONY), where MPPC stands for Multi-Pixel Photon Counter.
This system is characterized by the optical sensor it uses.
This sensor is a MPPC sensor customised for astronomical observations.
Each pixel consists of a Geiger APD and a quenching resistor, with all pixels connected in parallel,
effectively forming a monolithic Geiger APD array.
When a photon enters the sensor, a Geiger discharge occurs in the APD.
The discharge is suppressed by the voltage drop across the quenching resistor,
resulting in a pulse width of a few nanoseconds and a high signal-to-noise ratio sufficiently sensitive to a single photon.
The sensor comprises 64 pixels, arranged in an $8\times 8$ configuration.
Each pixel has a size of 75, 100, 150 or $200$\,$\mu$m square.
The maximum photon detection efficiency (PDE) is $\sim 70$\% at $480\,\mathrm{nm}$.
%This sensor enables independent readout for each pixel,
%is sensitive to a single photon, and hence provides a high signal-to-noise ratio.

\begin{figure}[thb]
\centering
\includegraphics[width=0.7\textwidth]{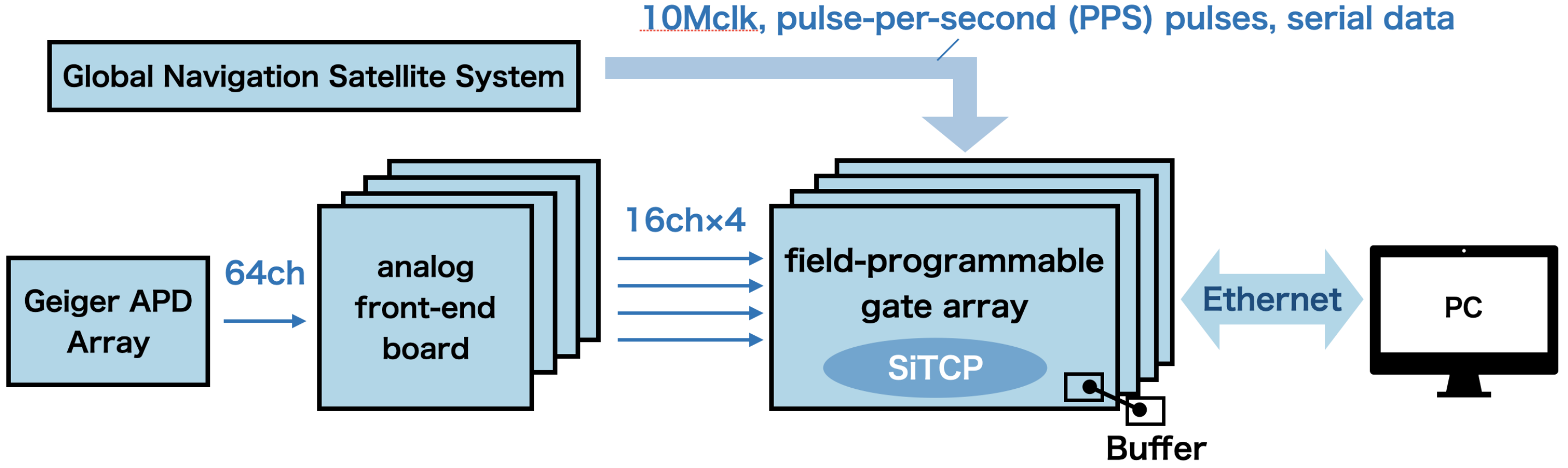}
\caption{Block diagram of the previous IMONY system.}
\label{IMONY16ch}
\end{figure}

Figure~\ref{IMONY16ch} shows a block diagram of the previous IMONY system, which assigns timestamp for each detected photon event by properly processing hit pixels and time information.
The 64-channel signals from the sensor are processed by four readout systems.
Each readout system handles 16 channel 
and consists of an analog front-end (AFE) board and a field-programmable gate array (FPGA).
First, in the AFE board, 
analog signals are amplified and converted into timing pulses based on a threshold voltage.
Then, FPGA receives the timing pulses and generates photon hit data 
using time information from the Global Navigation Satellite System (GNSS) receiver.
The processed data are stored in buffers and transmitted to a PC 
using SiTCP, data transfer technology \cite{sitcp}.

%The process of FPGA-based data generation is illustrated in Figure~\ref{datagen}.
The FPGA generates timestamped photon data during the observation run. At the start of the observation run, the GNSS time information is sent to the PC,
which provides the absolute time reference for the beginning of the run.
After the observation starts,
the FPGA continuously checks for photon hits every 5\,ns.
If a photon is detected in any pixel, 
the system generates a data sequence 
by combining the hit pixel information and the detection time. 
If no photon is detected, no data are generated.
The detection time of each photon is recorded using two counters.
One is the pulse-per-second (PPS) pulses counter and the other is the 10\,MHz clock counter.
The 10\,MHz counter resets every time a PPS signal is received.
The FPGA stores both counters in the generated data sequence,
allowing the reconstruction of the relative time of each photon detection
from the start of the observation run.
The timing jitter for the recorded time stamp was conservatively estimated less than 200\, ns \cite{Nak2}.

%\section{Observation result}
%\label{sec:obs}

We have previously conducted observations with IMONY
at the $3.8\,\mathrm{m}$ Seimei Telescope in Okayama, Japan.
These included stellar observations for performance evaluation \cite{hasebe}
and Crab pulsar observations \cite{Has, HasD}, both of which were successfully carried out.
However, there were some issues. 
One major issue was the large number of components in the system,
which resulted in significant time required for mounting the system onto the telescope.
Another issue was unstable board-to-board connections,
where poor contact between the wiring occasionally led to interruptions in observations.
Furthermore, the need to install and remove the system for each observation period
increased the risk of deterioration and damage.
To address these issues, we developed a new board
that integrates the readout system of IMONY into a unified structure.

\section{New system}
\label{sec:FGATI}

\begin{figure}[b]
\centering
\begin{minipage}[b]{0.35\columnwidth}
    \centering
    \includegraphics[width=\columnwidth]{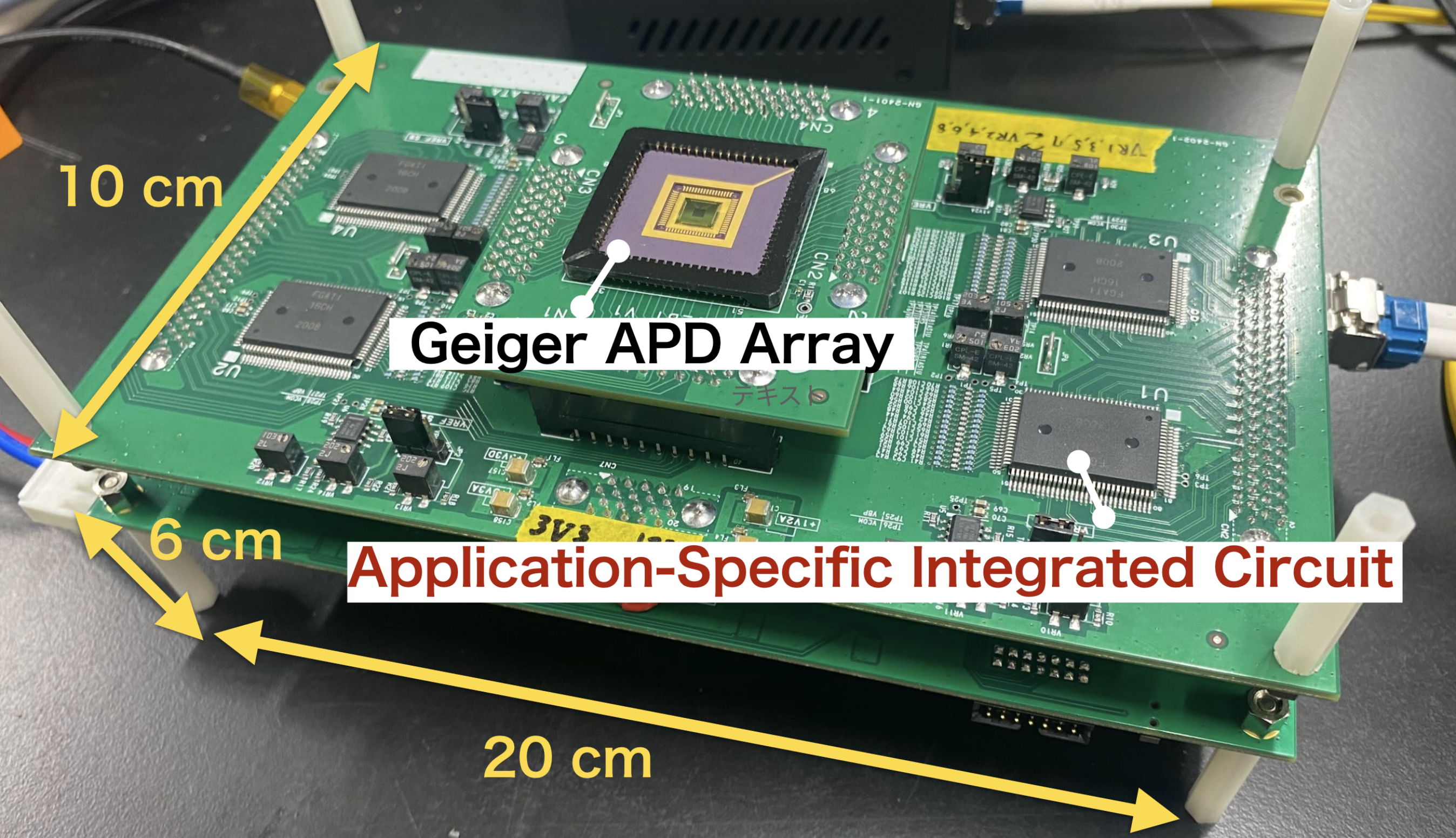}
    %\caption{Image of the new board system. It consists of three interconnected boards, the first of which includes the sensor.}
    \caption{Photograph of the new board system with three stacked boards.}
    \label{actualboard}
\end{minipage}
\hspace{0.03\columnwidth}
\begin{minipage}[b]{0.6\columnwidth}
    \centering
    \includegraphics[width=\columnwidth]{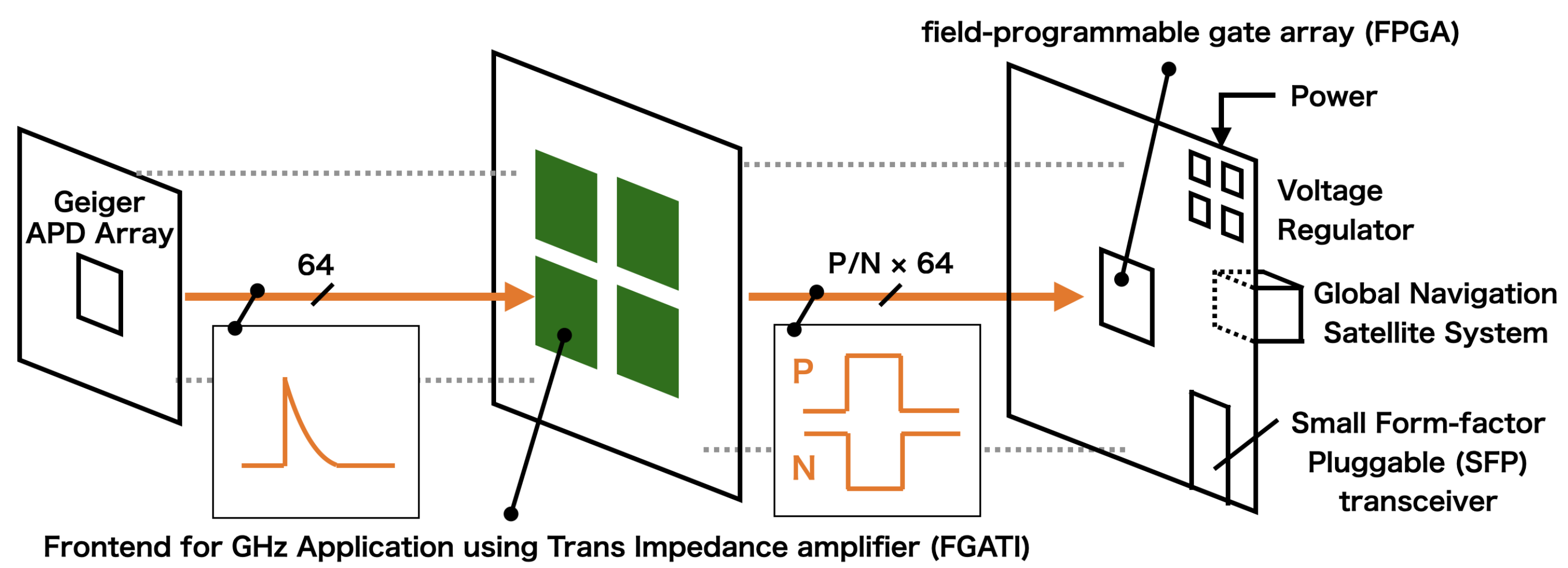}
    \caption{Schematic overview of the new board system and signal flow.}
    \label{overview}
\end{minipage}
\end{figure}

Although both the previous IMONY system (Figure~\ref{IMONY16ch}) and the newly developed version support 64 channels, the former was constructed from four independent 16-channel readout boards, each equipped with its own FPGA and Ethernet interface. This modular approach required multiple cables and synchronization between units.
In contrast, the new system integrates all 64 channels into a single board with one FPGA and a single Ethernet output, significantly improving compactness, reliability, and ease of deployment.
Figure~\ref{actualboard} shows an actual image of the new board system that we developed.
The board measures approximately 10\,cm in height, 20\,cm in width, and 6\,cm in overall thickness.
It consists of three interconnected boards, with the sensor mounted on the first board.
These boards are connected via board-to-board connectors, eliminating the need for detachable cables.
Schematic view and signal flows are shown in Figure\,\ref{overview}.

The sensor signals from the first board are transmitted in parallel 
through 64 channels to the second board.
The second board houses four Frontend for GHz Application using Trans Impedance amplifier (FGATI) chips, 
an analog Application-Specific Integrated Circuit (ASIC) \cite{fgati}. 
Each FGATI chip processes 16 channels.
In this system, IMONY utilizes the fast transimpedance amplifier within FGATI,
which amplifies the analog current input 
before being processed by the integrated comparator in FGATI.
The comparator applies a threshold to the signal 
and outputs a low-voltage differential signaling (LVDS) hit pulse.
The comparator threshold for each channel can be 
independently configured via SPI (Serial Peripheral Interface) communication.
The 64-channel LVDS signals are then transmitted to an FPGA located on the third board,
where hit pulse detection and timestamp generation take place.
The data generation algorithm remains unchanged from the current version.
Data transfer is still performed via Ethernet using SiTCP.
In this system, however, the output is transmitted via an optical fiber 
using a Small Form-factor Pluggable (SFP) module.
The system is powered by a single 5\,V input,
with all necessary voltages generated by regulators on the third board.

 \begin{figure}[htbp]
\centering
\includegraphics[width=0.7\textwidth]{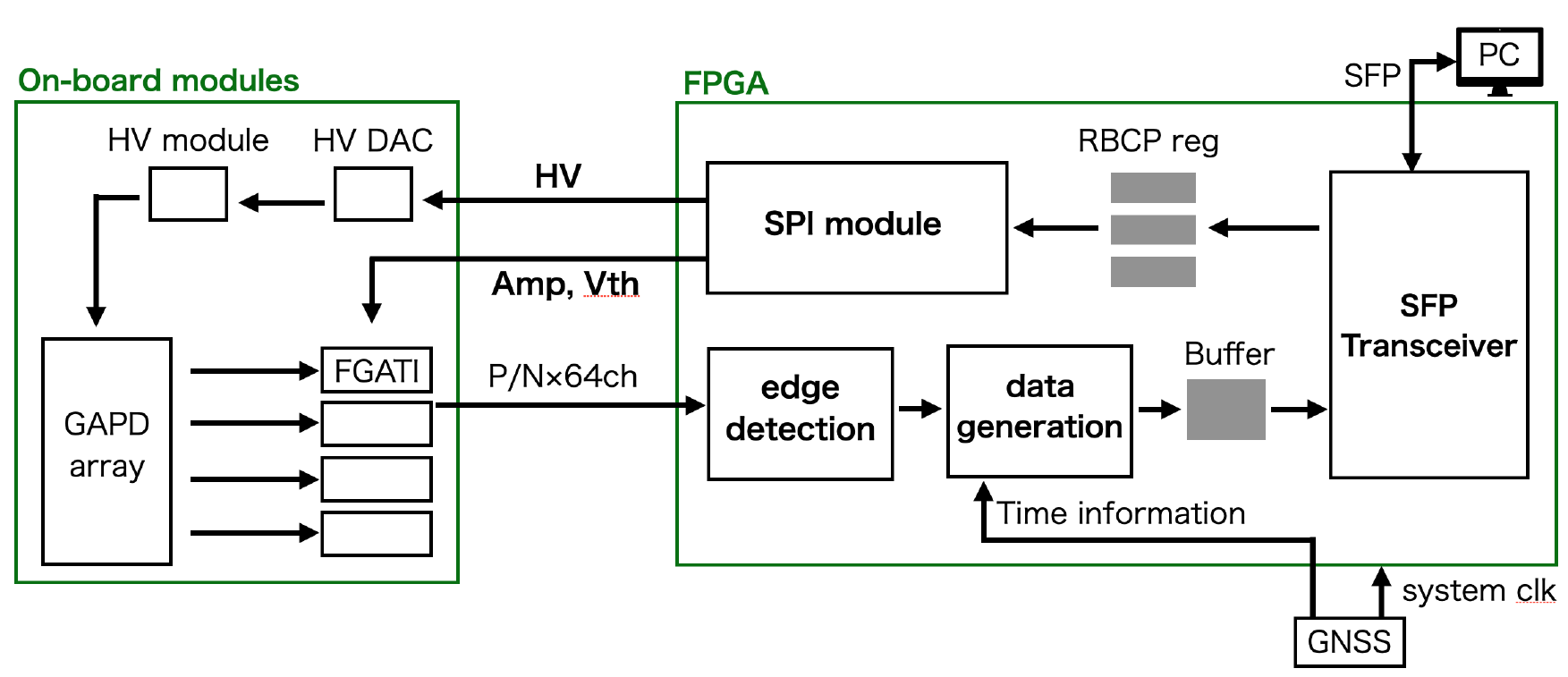}
\caption{Block diagram of FPGA operation and module control in the new system.}
\label{newconfig}
\end{figure}

We also implemented FPGA coding to control the system.
Figure~\ref{newconfig} illustrates the FPGA operation and its connections with the system modules.
First, each module on the board is controlled by commands from the FPGA.
The FPGA operates using the GNSS clock as the system clock.
The commands issued by the FPGA are generated based on parameters received from the PC via Ethernet.
In addition, the high voltage and the FGATI chips on the board are controlled via SPI.
The FPGA interprets the parameters stored in registers on the internal bus called Remote Bus Control Protocol (RBCP),
and converts them into SPI commands.
By adjusting the high voltage and threshold settings in FGATI chips,
the analog signals from the sensor are appropriately converted into timing pulses.
When the FPGA detects a photon hit, the hit pixel information and the photon arrival time
are combined into a data packet. 
The acquired data are stored in a fast-in-fast-out buffer in the FPGA and finally transmitted to the PC.

The timestamp resolution of the system is 100\,ns, determined by the 10\,MHz reference clock provided by the GNSS module (FURUNO GF-8803). According to the manufacturer's specifications, the timing jitter of the reference clock is on the order of tens of nanoseconds, which ensures reliable timing tagging for photon events. While the previous version of IMONY has been evaluated for dynamic range and timing jitter in earlier studies~\cite{hasebe,Nak2}, quantitative evaluation of the new integrated system is currently ongoing and will be reported in future work.

\section{Evaluation}

\begin{figure}[htb]
\centering
\begin{minipage}[b]{0.45\columnwidth}
    \centering
    \includegraphics[width=\columnwidth]{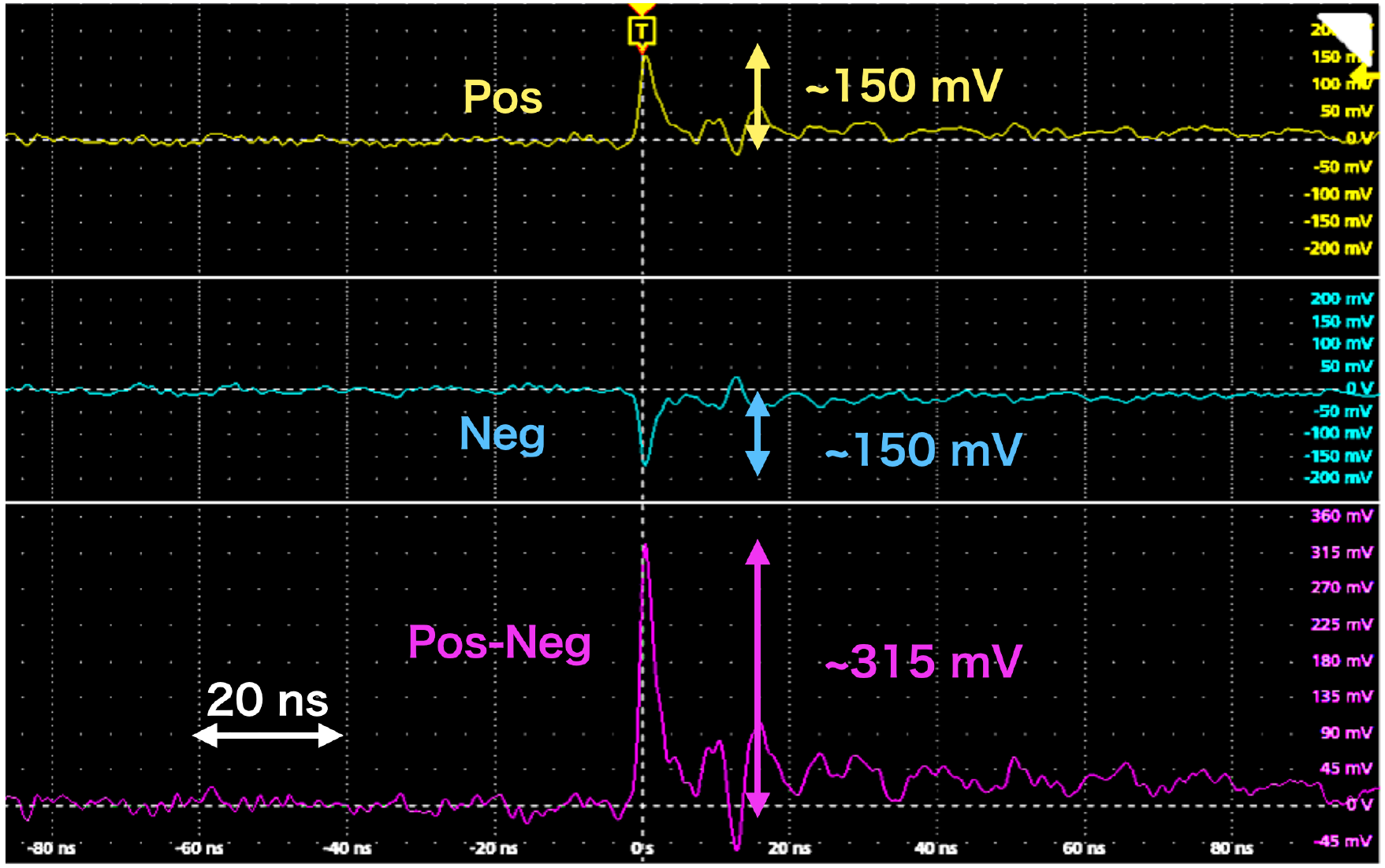}
    \caption{LVDS output from the IMONY sensor, triggered by a dark count.}
    \label{dout}
\end{minipage}
\hspace{0.03\columnwidth}
\begin{minipage}[b]{0.45\columnwidth}
    \centering
    \includegraphics[width=0.8\columnwidth]{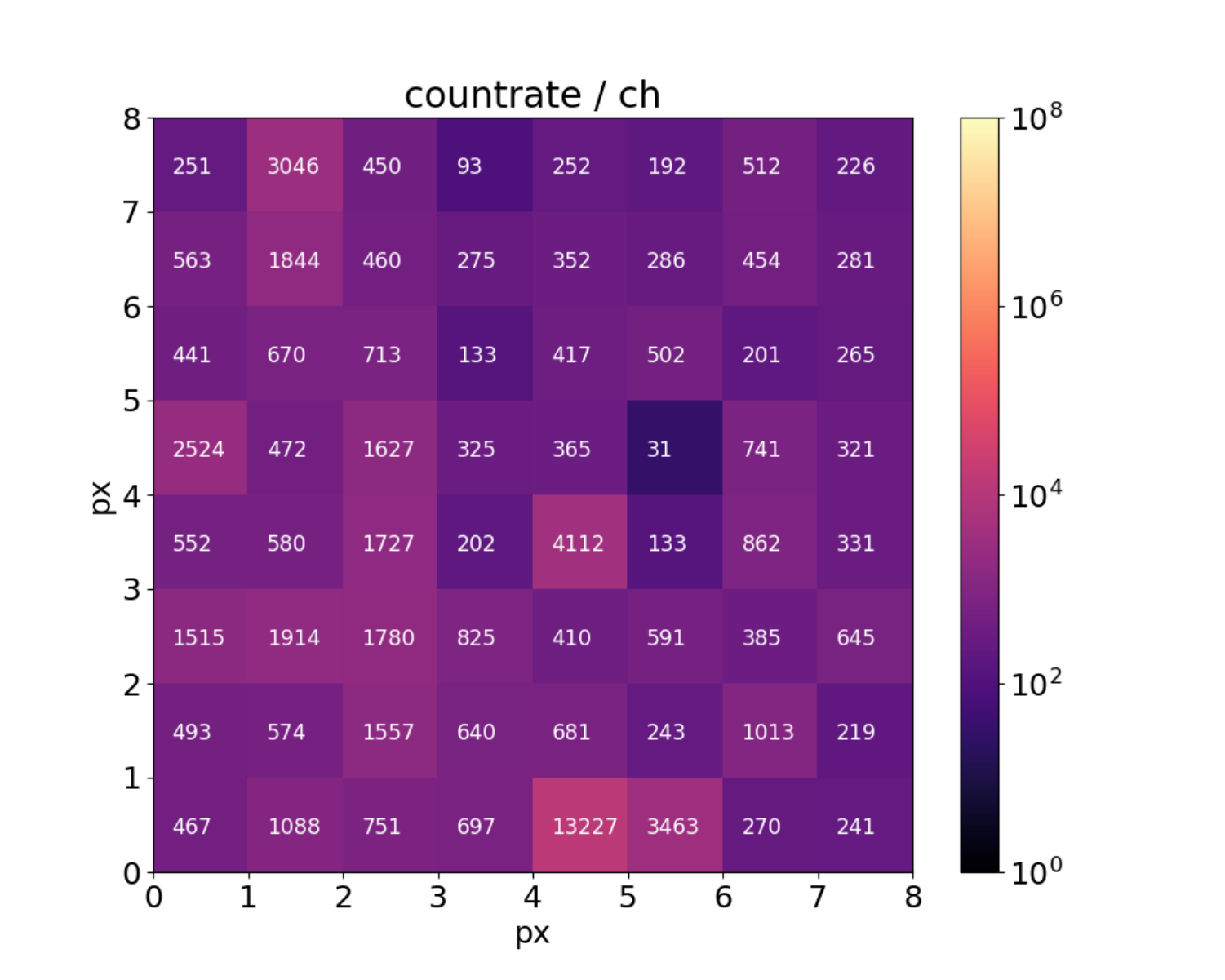}
    \caption{Dark count rate map for all 64 pixels, measured at room temperature of $25-6^\circ$C.}
    \label{darkmap}
\end{minipage}
\end{figure}

Figure~\ref{dout} shows an example of the digital output signals from an FGATI
after applying an appropriate threshold setting to the FGATI
and adjusting the high voltage to the sensor.
When the sensor was shielded from light, the LVDS standard differential outputs as shown in Figure~\ref{dout} were confirmed in all 64 channels.
Figure~\ref{darkmap} shows a 64-pixel count rate (white text) arranged in the sensor field of view at room temperature of $25-6^\circ$C.
This count rate was simultaneously observed across all 64 pixels using preliminary threshold settings that have not yet been fully optimized.
This result, shown in Figure~\ref{darkmap}, demonstrates that the pulse detection and time-stamping functions of all 64 channels are operating correctly under dark conditions. The uniform presence of dark counts across all pixels confirms that the sensor signals are successfully processed and recorded by the system.

\begin{figure}[htb]
\centering
    \includegraphics[width=0.5\columnwidth]{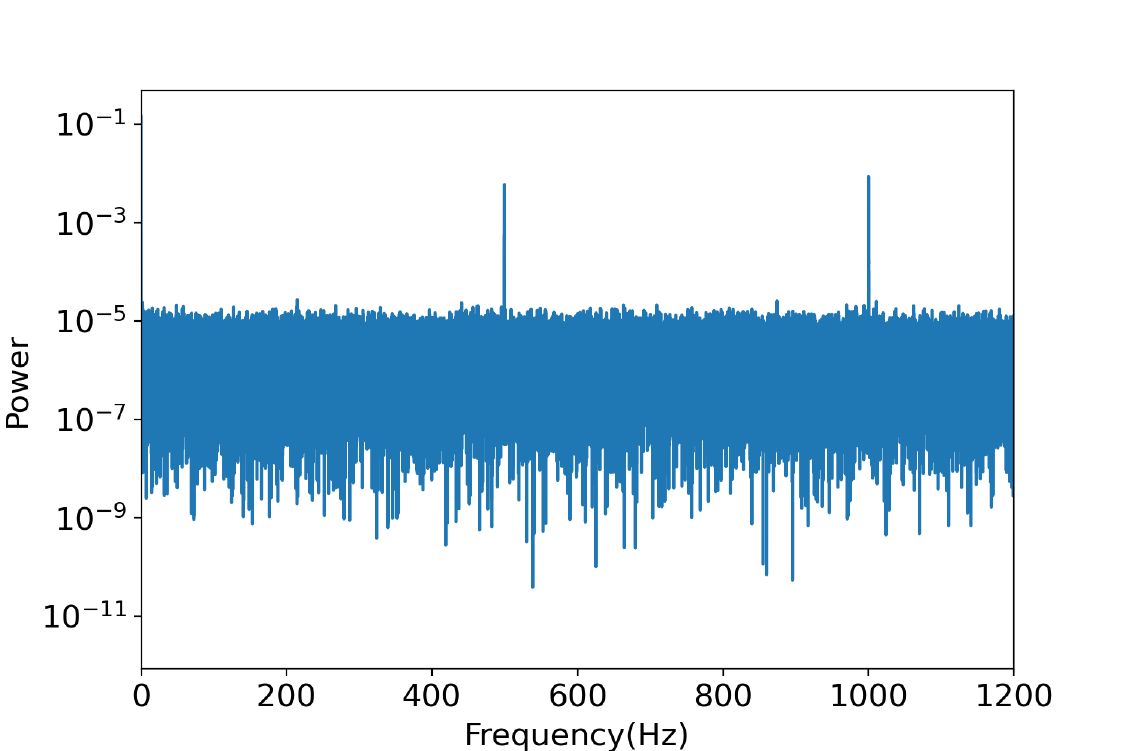}
     \caption{Example of the Fourier power spectrum for periodic picosecond laser illumination at 500 Hz.}
        \label{fft}
\end{figure}

We then conducted an experiment using a periodic light source in our laboratory.
A picosecond laser \cite{inome} was used as the periodic light source,
and illuminating the entire sensitive area uniformly.
The laser had a wavelength of 405\,nm and a pulse duration of approximately 80\,ps per shot.
The laser was operated using a trigger signal from a clock generator.
Figure~\ref{fft} presented an example of the Fourier power spectrum
when the laser was triggered at 500\,Hz.
Fourier transform was performed on all channels, and peaks were confirmed at 500\,Hz and its double frequency on each channel.
We confirmed that all channels successfully detected the periodic pulses
at the expected frequency and intend to evaluate the accuracy in the future.

\section{Summary and future work}
\label{sec:summary}
We have developed a photon-counting imager system IMONY,
and tested the newly developed readout boards employing FGATI, an analog ASIC.
We implemented FPGA coding, enabling the operation of each module,
and demonstrated in a laboratory setting that the system can perform precise light curve measurements.
Future work includes observational experiments with a telescope
to further evaluate and enhance the system’s performance.
The new board system has achieved higher integration and miniaturization of the entire system.
With multiple units of such systems installed on a telescope,
we have begun exploring applications for a high-speed, multi-color simultaneous imaging system.
Additionally, the system’s compactness has improved portability,
allowing for its integration into portable small telescopes for occultation observations. 
%Occultation is a phenomenon in which an asteroid temporarily shades out the light of a star
%as it passes in front of it and is observed as a transient variation in light intensity.
Occultation is a phenomenon in which an asteroid temporarily blocks the light from a star by passing along the line of sight, resulting in a transient dip in brightness.
Although not optimized for dense tiling, the system allows for possible expansion using larger-area sensors or synchronized modules.
With further development, this system will contribute to a better understanding of various fast fluctuation phenomena of light intensity.

\acknowledgments

This work was supported by JSPS KAKENHI Grant Numbers 23H01194.
We also thank the anonymous referee for giving helpful comments to improve the manuscript.

% Bibliography

\end{document}